\documentclass[conference]{IEEEtran}
\IEEEoverridecommandlockouts

\usepackage[noadjust]{cite}

\usepackage{epstopdf}

\ifCLASSINFOpdf
\usepackage[pdftex]{graphicx}
\else
\usepackage[dvips]{graphicx}
\fi

\usepackage{multicol, blindtext}
\usepackage{amsmath}

\usepackage{array}
\usepackage{calc}

\begin{document}
	
	\title{FPGA Implementation of High Speed Reconfigurable Filter Bank for Multi-standard Wireless Communication Receivers}

	
	%
	\author{\IEEEauthorblockN{Sasha Garg and S. J. Darak}
		\IEEEauthorblockA{Department of Electronics and Communication Engineering\\
			Indraprastha Institute of Information Technology (IIIT), Delhi, India 110020\\ 
			Email: \{sasha15109, sumit\}@iiitd.ac.in}}


	\IEEEpubid{\begin{minipage}{\textwidth}\
			\textbf{978--1--5090--1422--4/16/\$31.00~\copyright~2016 IEEE}\\
		\end{minipage}}
		\maketitle

	\begin{abstract}
		In next generation wireless communication system, wireless transceivers should be able to handle wideband input signals compromising of multiple communication standards.Such multi-standard wireless communication receivers (MWCRs) need filter bank to extract the desired signal of interest from wideband input spectrum and bring it to the baseband for further signal processing tasks such as spectrum sensing, modulation classification,demodulation etc.In MWCRs,rather any wireless receivers, modulated filter banks, such as Discrete Fourier Transform Filter Banks (DFTFB), are preferred due to their advantages such as lower area, delay and power requirements. To support multi-standard operation, reconfigurable DFTFB (RDFTFB) was proposed by integrating DFTFB with the coefficient decimation method. In this paper, an efficient high speed implementation of RDFTFB on Virtex-7 field programmable gate arrays (FPGA) has been proposed.The proposed approach minimizes the critical path delay between clocked registers thereby leading to significant improvement in the maximum operating frequency of the RDFTFB. Numerically, the proposed implementation leads to 89.7\% improvement in the maximum frequency at which RDFTFB can be clocked.Furthermore,proposed implementation leads to 18.5\% reduction in the dynamic power consumption.
	\end{abstract}
	

	\IEEEpeerreviewmaketitle
	
	     \IEEEpubidadjcol
	
	\section{Introduction}
	Paradigms such as opportunistic spectrum access based cognitive radio, unlicensed LTE and the need of wide variety of services from data to multimedia results in wideband input signals compromising of multiple communication standards with different specifications~\cite{hentschel1999digital,hentschel1999spectrum,pucker2003channelization,farhang2008filter}.
	The first task of any wireless receiver is to choose the desired signal of interests and shift it to the baseband followed by conventional signal processing tasks such as spectrum sensing, modulation classiﬁcation, demodulation etc. Conventional receivers, which are designed to process input signal consisting of only one communication standard, prefer modulated uniform filter bank such as Discrete Fourier Transform Filter Banks (DFTFB), Fast Filter Bank (FFB) etc. due to their lower area, delay and power requirements. However, in multi-standard environment, a reconﬁgurable ﬁlter bank with tunable subband bandwidth is desired. Such receivers are referred to as Multi-standard wireless communication receivers (MWCRs). Furthermore, in addition to lower area and power requirements, such reconﬁgurable ﬁlter banks should be fast enough to sense dynamically changing environment i.e., spectrum. \par Various techniques have been proposed in the past for channelization in MWCRs. One of the first and straightforward approach is Per-Channel (PC) approach~\cite{hentschel2002channelization}. PC approach consists of parallel banks of digital filters, digital down converters and digital down samplers, one for each received subband. Though PC approach offers multi-standard channelization, implementation complexity linearly increases with the number of received subbands. Another approach, known as Pipelined Frequency Transform (PFT)~\cite{lillington} consists of a binary tree of digital down converters and sampling rate converters. Its complexity is lower than the PC approach. Since, the signal is split into two halfbands at each stage, the sampling rate and hence, dynamic power consumption also decreases at every stage. The modulated filter banks~\cite{hentschel2002channelization,mahesh2010reconfigurable1} provide low complexity alternative to PC and PFT approaches. For instance, Discrete Fourier Transform Filter Bank (DFTFB) is a modulated uniform filter bank consisting of prototype lowpass filter followed by DFT modulator. Due to their low complexity, DFTFB are widely employed in wireless communication receivers. However, DFTFB is a uniform filter bank and hence, a separate DFTFB would be required for each received communication standard in MWCR, which leads to huge area complexity. Frequency response masking (FRM) based uniform and non-uniform filter banks for MWCRs have been proposed in~\cite{darak2010reconfigurable,darakffb}. Such filter banks have sharp transition bandwidth but the group delay is very high. In \cite{mahesh2008coefficient}, a reconfigurable DFTFB (RDFTFB) was proposed by integrating DFTFB with the coefficient decimation method (CDM). RDFTFB offers tunable subband bandwidth using fixed-coefficient lowpass prototype filter. The implementation complexity of RDFTFB is slightly higher than that of DFTFB. The filter bank in \cite{darakspa} is designed using spectral parameter approximation based variable digital filter and Farrow structure. It offers complete control over the bandwidth as well as center frequency of each subband. Also, the group delay is very low. However, the overall complexity is huge compared to DFTFB and RDFTFB.
	
	\par In this paper, an efficient high speed implementation of RDFTFB on Virtex-7 field programmable gate arrays (FPGA) has been proposed. The proposed approach minimizes the critical path delay between clocked registers in the prototype lowpass filter as well as DFT modulator block of the RDFTFB. This leads to significant improvement in the maximum operating frequency of the RDFTFB. Numerically, the proposed implementation leads to 89.7\% improvement in the maximum frequency at which RDFTFB can be clocked. Furthermore, proposed implementation leads to 18.5\% reduction in the dynamic power consumption. 
	\par The paper is organized as per the following. In section II, the design of RDFTFB is discussed in detail followed by FPGA implementation of RDFTFB in Section III. In Section IV, proposed high speed implementation of RDFTFB is discussed. In section V, implementation results are presented followed by conclusions in Section VI.

	\section{RDFTFB Design}
	In this section, design details of RDFTFB in~\cite{mahesh2010reconfigurable1} are discussed.
	\subsection{DFTFB}
	DFTFB is a uniform filter bank consisting of the prototype lowpass filter of length $L$ and impulse response, $h(n)$ where $ n= {1,2,..L}$. The passband ripple, stopband attenuation and transition bandwidth of the prototype filter are same as the desired specifications of the DFTFB, $\delta_p$, $\delta_s$ and $\Delta$, respectively. For $N$-subband DFTFB, the bandwidth of the prototype filter is equal to $(1/N)$. Note that all the frequency specifications mentioned hereafter are normalized with respect to half of the sampling frequency. Then, frequency response of any $k^{th}$ subband of the DFTFB is given by~\cite{hentschel2002channelization},
	
	\begin{equation} 
	y_{k}[n]=\sum\limits_{i=0}^{N-1}h[n-i]e^{j\frac{2\pi}{N}ki}
	\end{equation}
	where $h[n]$ is the response of the prototype filter and $y_{k}[n]$ is the response of the $k^{th}$ subband. For efficient implementation, prototype lowpass filter is generally implemented in transposed polyphase decomposed form. It is apparent from Eq. (1) that the $N$ subbands provided by the conventional DFTFB are of uniform and fixed bandwidth. In order to support multi-standard channelization in MWCRs, subband bandwidth needs to be made tunable. The design of such reconfigurable filter bank is discussed in the next subsection.

	\subsection{Reconfigurable DFTFB (RDFTFB)}
	The straightforward approach to change the subband bandwidth would be to use programmable prototype filter in which filter coefficients are updated every time subband bandwidth needs to be changed. This approach incurs huge reconfiguration delay and hence, may not be suitable for MWCRs which operates in dynamically changing spectrum environment. Another approach is the use of coefficient decimation method (CDM) to make the bandwidth of lowpass prototype filter, $h(n)$ tunable without the need of coefficient updates. Such filter bank is referred to as RDFTFB~\cite{mahesh2008coefficient} and is shown in Fig. 1.
	\begin{figure}[htbp]
		\centering  \includegraphics[width=1.00\linewidth]{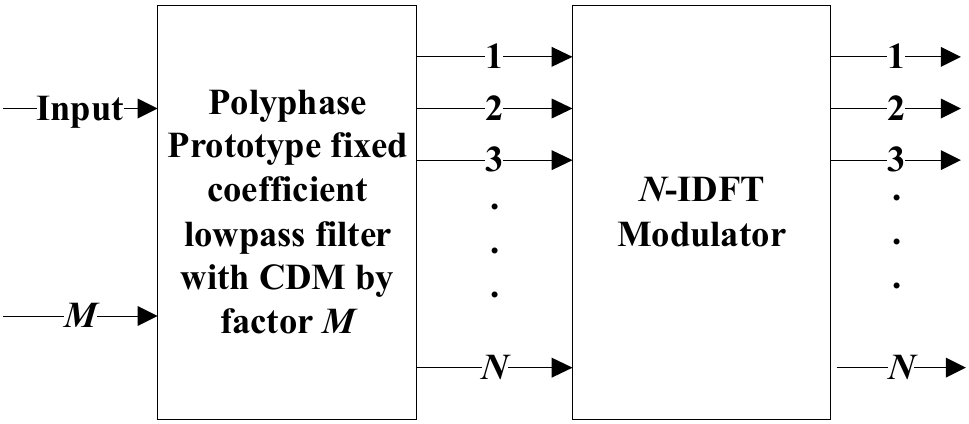}
		\caption{Block Diagram of the implemented system}
		\label{fig:1}
	\end{figure}
	
	The RDFTFB uses CDM (referred to as CDM-II in~\cite{mahesh2010reconfigurable1}). In CDM by factor $M$, every $M^{th}$ coefficient is retained. Then, all the retained coefficients are grouped together. For example, consider a filter, $h(n)$, with coefficients as ${h_1, h_2, h_3, h_4,...,h_L}$. Then, the coefficients of the resultant decimation filter, $h'(n)$ after CDM by factor $M=2$, is given by ${h_1, h_3, h_5,...,h_L}$. For illustration, the frequency responses of $h'(n)$ for $M={1,2,3,4,5}$ are shown in Fig. 2. It can be observed that bandwidth of the prototype filter can be changed by changing the value of $M$. Note that as $M$ increases, stopband attenuation as well as transition bandwidth becomes worse. This can be overcome by using higher order prototype filter. 
	To avoid aliasing, the value of $M$ has to be limited in a manner such that,
	
	\begin{equation} 
	M*f_{o}<\pi
	\end{equation}
	
	Where $f_{o}$ is the bandwidth of the prototype filter $h(n)$.
	\par In RDFTFB, the subband bandwidth is changed by controlling the decimation factor $M$ as shown in Fig. 1. Thus the bandwidth of each RDFTFB subband is $M$ times the bandwidth of the prototype filter. 
	\begin{figure}[htbp]
		\centering
		\includegraphics[width=\linewidth,scale=2]{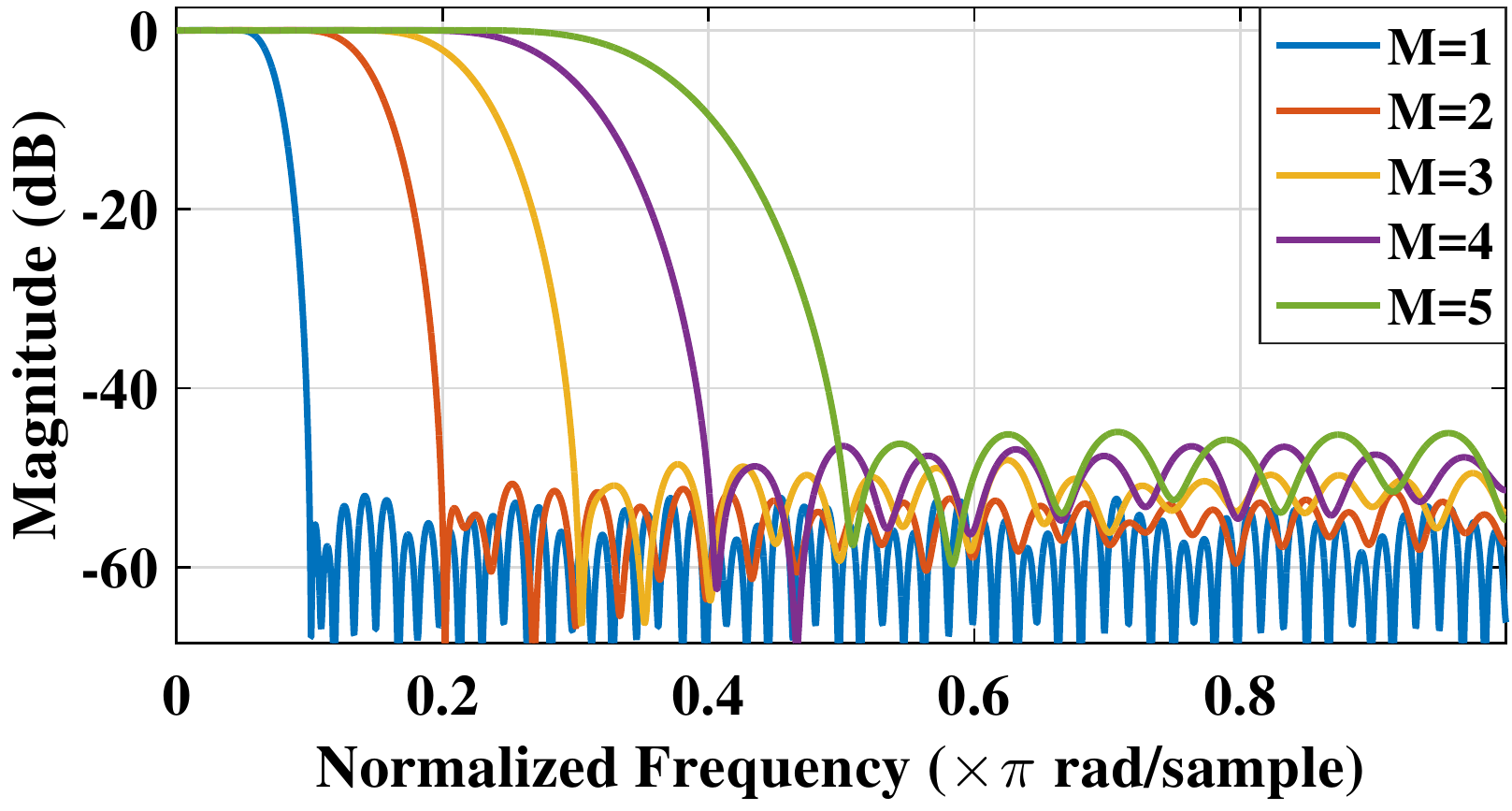}
		\caption{Filter responses for different values of M (M=1,2,3,4,5)}
	\end{figure}

	\section{Proposed High-Speed Implementation of RDFTFB}
	In this section, high-speed implementation
	of RDFTFB is discussed. First, the RDFTFB architecture is designed by integrating the CDM method with the prototype filter of DFTFB. Then, the critical path delay of the RDFTFB is reduced by adding clocked registers at suitably chosen locations in the architecture of the prototype filter as well as IDFT modulator. 
	
	\begin{figure*}[ht]
		\centering
		\includegraphics[width=\textwidth]{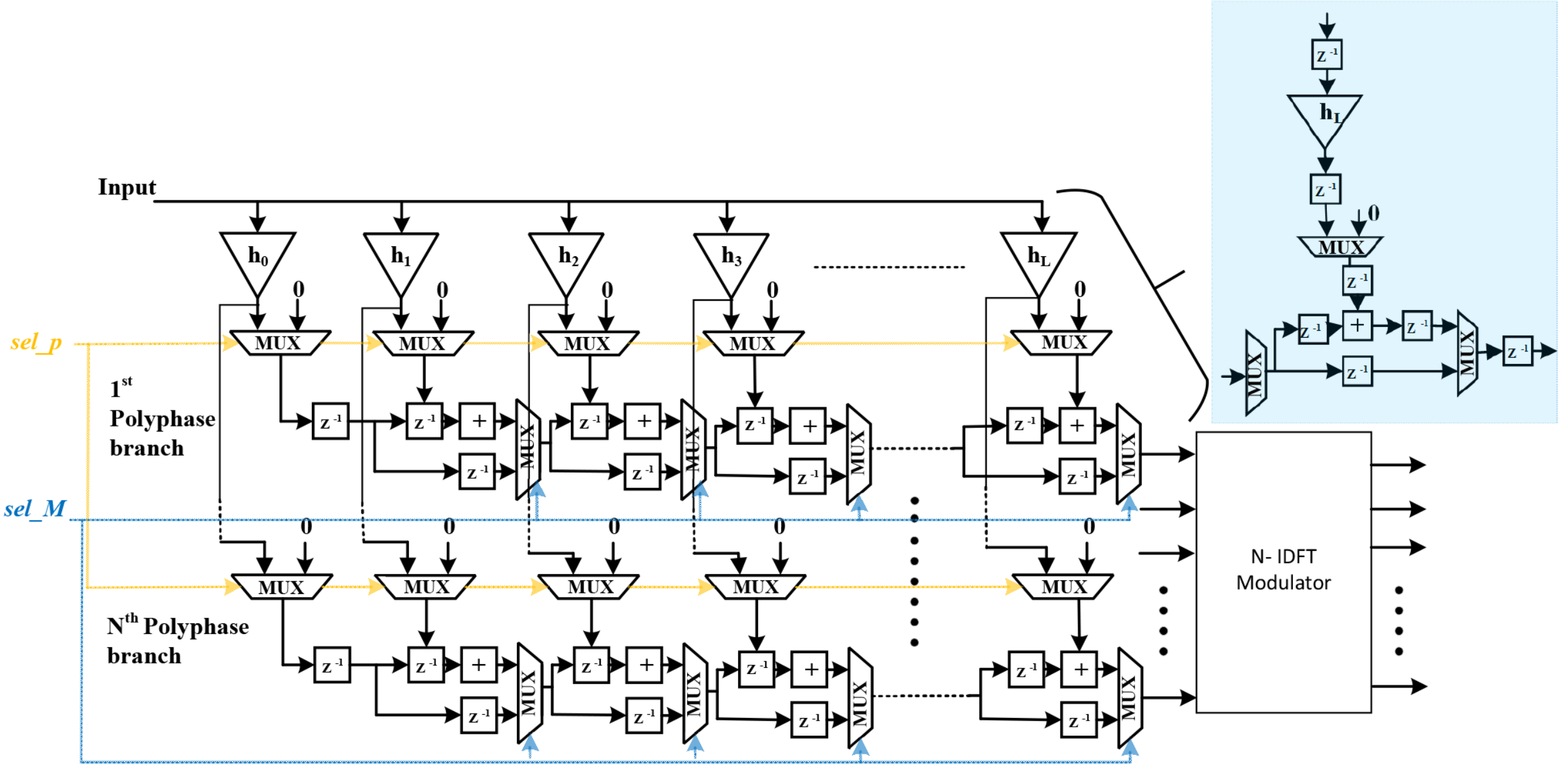}
		\caption{Architecture of RDFTFB}
	\end{figure*}
	
	\subsection{FPGA Implementation of RDFTFB}
	As shown in Fig. 1, the RDFTFB consists of prototype lowpass filter with CDM by factor $M$ and $N$-IDFT modulator. The detailed architecture of the RDFTFB is shown in Fig. 3. The prototype filter coefficients, $h_1, h_2,..,h_L$, for a given frequency specifications are obtained using $firpm$ command of MATLAB. The filter is implemented in transposed polyphase decomposed form with $N$ polyphase branches which are given by,
	\begin{equation} 
	H(z)=\sum\limits_{i=0}^{N-1} z^{-i}E_{i}(z^{N})
	\end{equation}
	where $ E_{i}'$s are the polyphase branches. The polyphase system is then followed by N-IDFT modulator as shown in Fig. 3. The multiplexer controlled signal, $sel_p$ is used to control appropriate filter coefficients for polyphase implementation of prototype filter while another signal, $sel_M$, decides the CDM factor, $M$. The output of each polyphase branch is then delayed by appropriate values and passed to the corresponding branches of IDFT modulator. The IDFT modulator consists of $N$ branches and is implemented using Eq. 1. The $k^{th}$ output of IDFT modulator consists of subband with center frequency of of $2\pi k/N, k=\{0,1,. . . ,N-1\}$ and bandwidth equal to $M$ times the bandwidth of the prototype filter. The RDFTFB needs $N_a$ number of extra adders and $2N$ 2-input multiplexers compared to the traditional DFTFB architecture where $N_a$ is given by:
	\begin{equation} 
	N_a=\sum\limits_{i=1}^M\Big(\frac{L}{i}-1\Big)
	\end{equation}
	
	Note that the prototype lowpass filter has linear phase and hence, its coefficients are symmetric. This means that the number of multipliers required to implement the prototype filter is half of the filter length, $L$. Next, high speed RDFTFB architecture to improve maximum operating frequency is discussed.

	\subsection{High Speed RDFTFB Implementation}
	In any combinational circuit, the signal travels through various logic paths. The signal experiences routing and logic delays while propagating from one clocked register to the other. Routing delay is the delay encountered by the signal due to the time it takes to travel along the wires and between the circuit elements in the architecture being implemented. Logic delay is the time taken by the signal to pass through various logic elements present between the clocked registers. The longest path that a signal travels between any two clocked registers defines the critical path of the architecture and corresponding delay is known as critical path delay, $\tau_{CPD}$. Thus, critical path delay can be characterized as the sum of the routing delay and the logic delay. However, it is dominated by the logic delays as the routing delays are comparatively negligible and generally optimized by the design tools. The maximum clock frequency, $f_{clk_{max}}$ at which the architecture can operate is inversely proportional to $\tau_{CPD}$.
	\begin{equation} 
	f_{clk_{max}}=1/\tau_{CPD}
	\end{equation}
	In practice, $\tau_{CPD}$ must be shorter than one clock period, $T_{clk}$, in order to guarantee the correct operation of the circuit. Then,
	\begin{equation} 
	T_{clk}>\tau_{CPD}+t_{setup}+t_{hold}
	\end{equation}
	where $t_{setup}$ and $t_{hold}$ are the setup and hold time of the clocked registers. 
	\begin{table*}[ht]
		\renewcommand{\arraystretch}{2.8}
		\caption{FPGA Implementation Results}
		\resizebox{\textwidth}{!}{
			\begin{tabular}{|p{8cm}|>{\centering\arraybackslash}p{4cm}|>{\centering\arraybackslash}p{5cm}|>{\centering\arraybackslash}p{5cm}|}
				\hline
				\textbf{Architecture} & \textbf{Slice Requirements} & \textbf{Dynamic Power Consumption (in W)} & \textbf{Maximum Frequency(in MHz)}\\
				\hline
				\textbf{Conventional RDFTFB\cite{hentschel2002channelization}} & 1645 & 0.264 & 59.5\\
				\hline
				\textbf{RDFTFB with reduced critical path delay of prototype filter} & 1620 & 0.254 & 70.5\\
				\hline
				\textbf{High speed RDFTFB} & 1925 & 0.215 & 112.9\\
				\hline
			\end{tabular}}
		\end{table*}
		
		In the proposed RDFTFB implementation, $\tau_{CPD}$ is reduced by inserting the clocked registers at appropriate locations.  For illustrations, the critical path of the branch containing filter coefficient, $h_L$, is reduced by inserting the clocked registers as shown using top-right highlighted box in Fig. 3. It can be observed that,critical path in the modified architecture consists of only multiplier compared to the multiplier, multiplexer and adder in case of original RDFTFB. Note that while adding the clocked registers, precaution needs to be taken such that functionality of the architecture remains the same. This is achieved by inserting additional clocked registers in the other branches of the architecture so that group delay of all the outputs is identical. In practice, $\tau_{CPD}$ of the prototype filter is relatively small due to delay elements as compared to $\tau_{CPD}$ of the IDFT modulator. For instance, in case of 8-subband RDFTFB, critical path delay of IDFT modulator consists of a multipliers and 7 adders compared to the multiplier, multiplexer and adder in case of prototype lowpass filter. Thus, inserting the clocked registers in just prototype filter architecture does not lead to significant improvement in maximum operating frequency as shown later in Section IV.In the proposed RDFTFB, clocked registers are also inserted in all the branches of IDFT modulator and necessary precaution is taken to maintain the functionality of the IDFT modulator. Next, the implementation results are discussed.
		
		\section{Implementation Results}
		In this section, implementation results of the 8-subband RDFTFB on Virtex-7 FPGA are presented. The normalized $\Delta$, $\delta_p$ and $\delta_s$ of the prototype lowpass filter are 0.1 and 0.04dB, 50dB respectively. For CDM, the decimation factor $M$ can be varied from 1 to 5. For these specifications, tunable responses for second and seventh subbands are shown in Fig. 7. and Fig.8. respectively. It can be observed that subband bandwidth can be controlled using $M$. Empirically, there is no effect on filter bank specifications such as tranisition bandwidth, passband ripple and stopband attenuation as well as functionality of the filter bank due to use of clocked registers.
		
		\begin{figure}[htbp]
			\centering
			\includegraphics[width=1.00\linewidth]{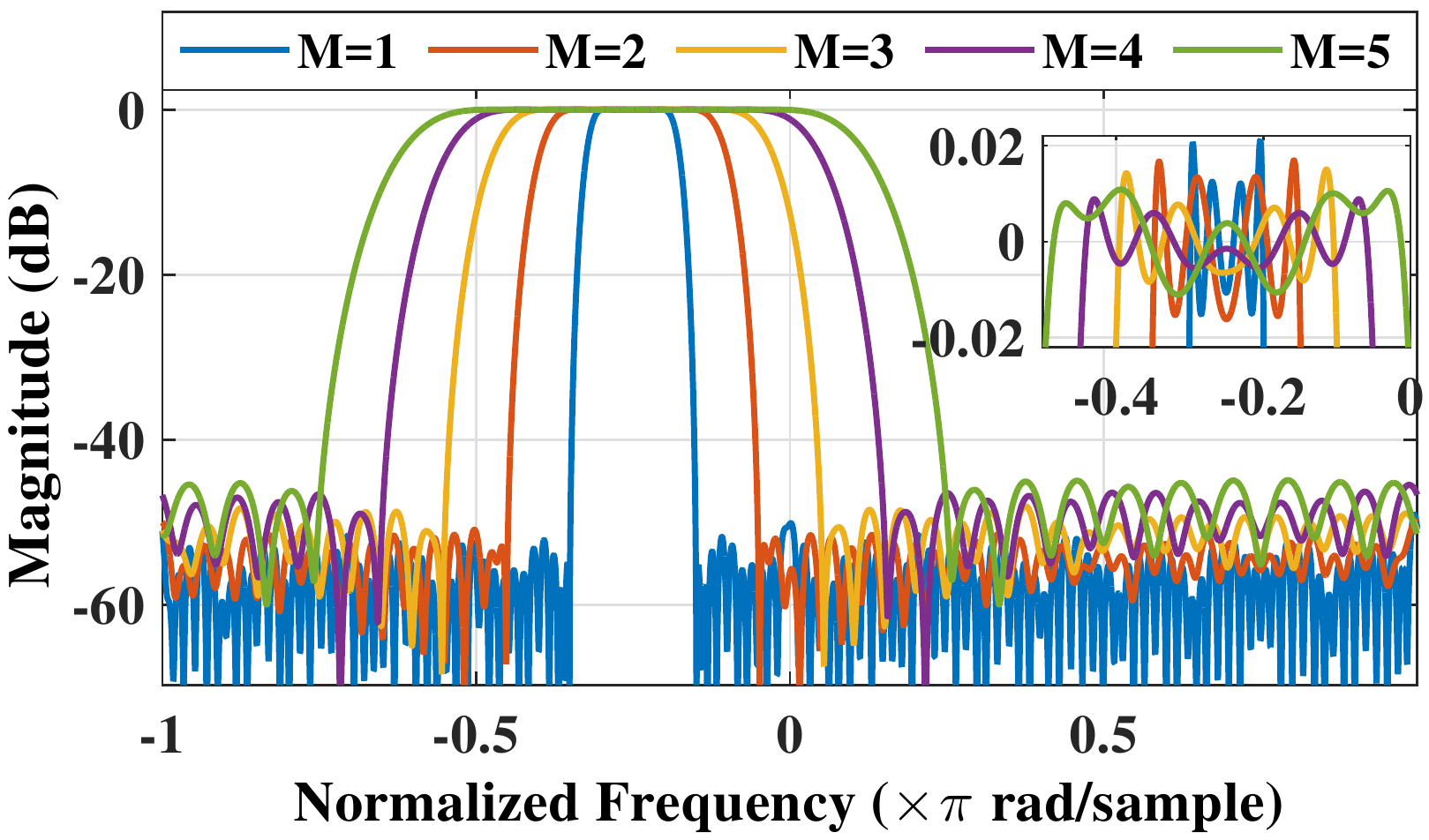}
			\caption{Frequency responses of the second subband of proposed RDFTFB for different values of $M$.}
		\end{figure}
		\begin{figure}[htbp]
			\centering
			\includegraphics[width=1.00\linewidth]{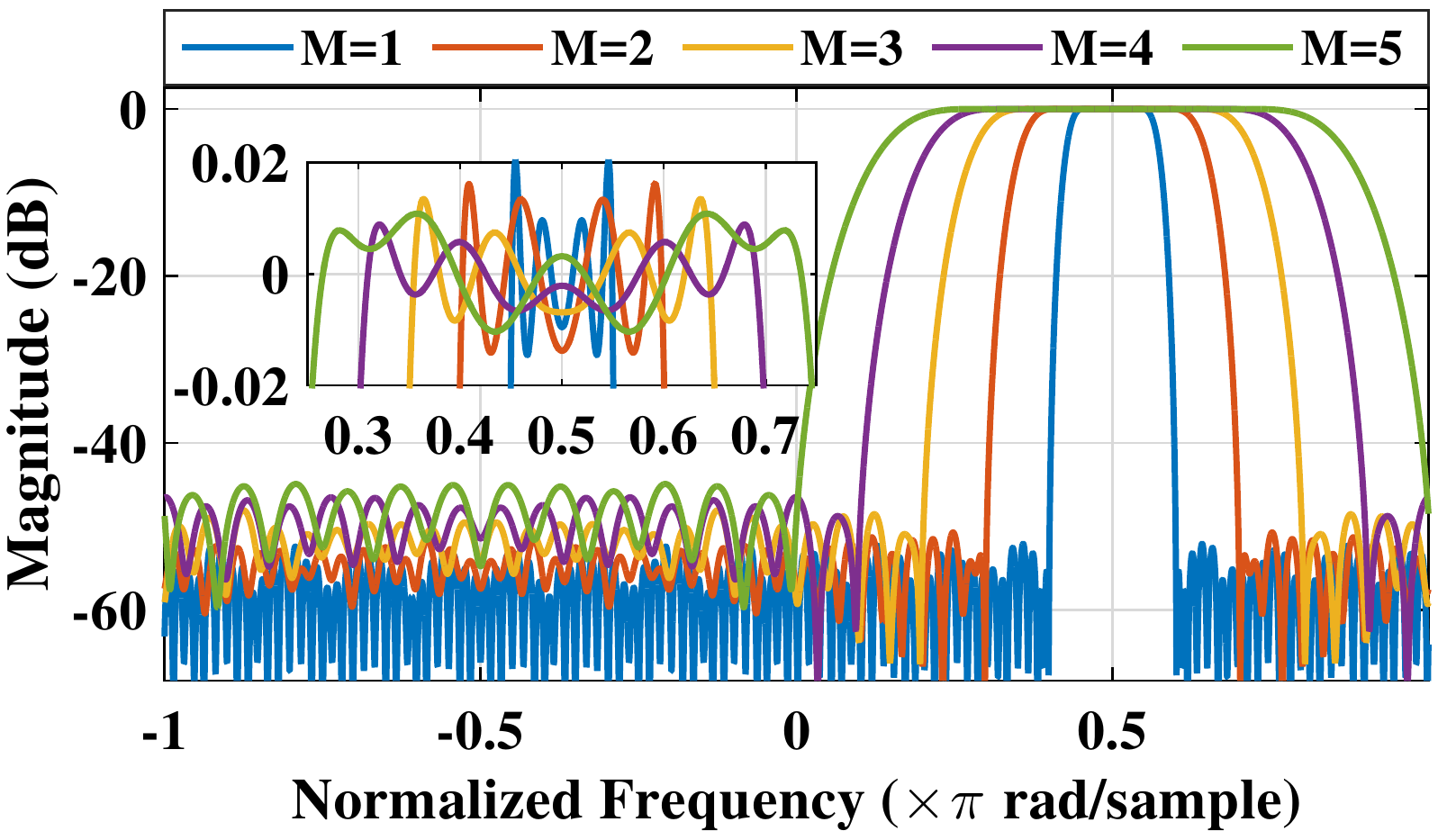}
			\caption{Frequency responses of the seventh subband of proposed RDFTFB for different values of $M$.}
		\end{figure}
		
		Next, the slice utilization, dynamic power consumption and maximum operating frequency of conventional RDFTFB, RDFTFB with reduced critical path delay of prototype filter and high speed RDFTFB (with reduced critical path delay of prototype filter as well as IDFT modulator) and high speed RDFTFB are compared as shown in Table. 1. It can be observed that proposed RDFTFB implementation leads to 89.7\% improvement in the operating frequency at which RDFTFB can be clocked. This is significant improvement considering conventional method of minimizing critical path delay of prototype filter leads to only 18.5\% in operating frequency. This proves that the proposed architecture has higher implementation speed. Furthermore, proposed implementation leads to significant savings of 18.5\% in the dynamic power consumption compared to the RDFTFB at respective maximum operating frequency. This means that, if conventional RDFTFB is operated at 112.9 MHz, then the total savings in dynamic power consumption would be much more than 18.5\% since higher the operating frequency, higher is the dynamic power consumption. The slice requirements of the proposed implementation is higher than conventional RDFTFB by 17\%. This is because of the need of clocked registers which in turn leads to higher number of slices. Note that total slice utilization of the proposed implementation is still less than 1\% of available slices on Virtex-7 FPGA. Thus, slight increase in slice utilization of proposed RDFTFB should not affect slice requirements other functions of MWCRs.

		\section{Conclusions and Future Work}
		In this paper, an efficient FPGA implementation of reconfigurable discrete Fourier Transform Filter bank (RDFTFB) has been presented. The proposed architecture leads to 89.7\% improvement in the maximum operating frequency of RDFTFB and 18.5\% reduction in total dynamic power consumption compared to the conventional RDFTFB. Future work involves further improvement in operating frequency and dynamic power consumption of RDFTFB using partial reconfiguration feature of FPGA.

		\section*{Acknowledgment}
		The authors would like to thank Department of Science and Technology (DST), Government of India for INSPIRE fellowship and Fund for Improvement of S\&T infrastructure in universities \& higher educational institutions (FIST) in support of this work.
		
			\bibliographystyle{IEEEtr}
			\bibliography{ref}
	\end{document}